\documentclass[12pt, a4paper]{article}
\usepackage[utf8]{inputenc}
\usepackage{lmodern}
\usepackage{tgschola}
\usepackage{titlesec}
\usepackage{amsmath}
\usepackage{graphicx}
\graphicspath{ {./images/} }
\usepackage{blindtext}
\usepackage{ifplatform}

\begin{document}
\begin{titlepage}
	\centering
	\includegraphics[width=0.7\textwidth]{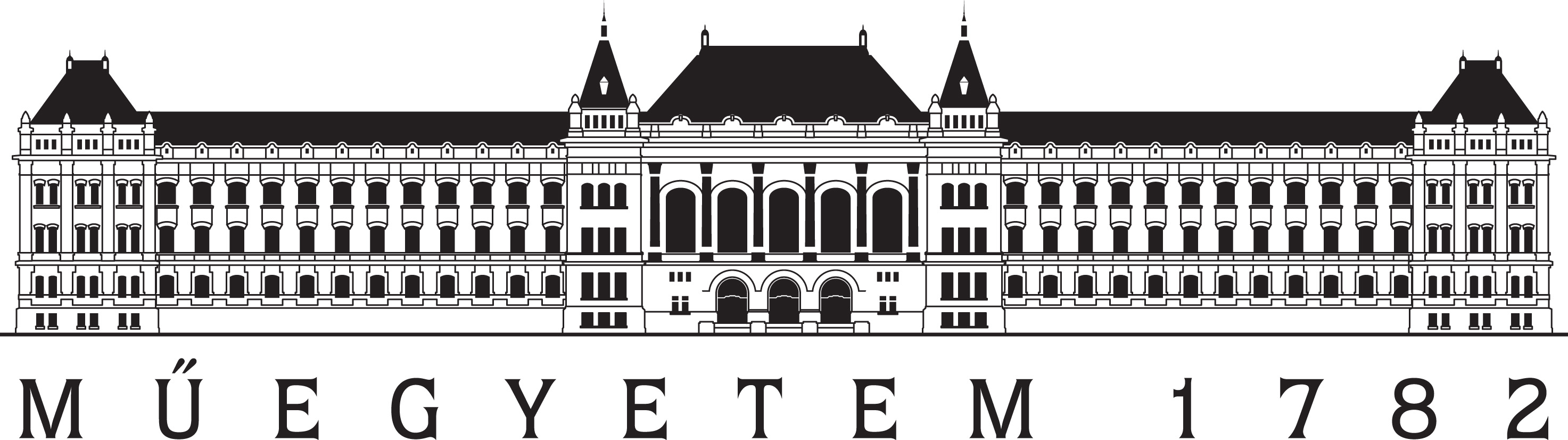}\par
	{\scshape Budapest University of Technology and Economics\par}
	{\scshape Faculty of Natural Sciences\par}
	{\scshape Institute of Mathematics\par}
	\vspace{1cm}
	{\huge The new science of COVID-19: A Bibliographic and Network Analysis\par}
	\vspace{1cm}
	{\Large Xuezhou Fan\par}
	{\large Budapest, Hungary}\\
	{\large 2021}
\end{titlepage}
\tableofcontents
\clearpage
\section{Introduction}
Corona viruses are a group of related RNA viruses that cause respiratory tract infections \cite{corona}. It includes COVID-19, SARS, MERS, etc. which claim many lives around the world. Since the outbreak of the COVID-19, the society has changed significantly, in order to restore the normal day-to-day life, more and more medical researchers have put the research focus on COVID-19, in a bid to find a solution to the pandemic. But later on COVID-19 related research emerged in a wide range of disciplines since the pandemic affected almost all segments of science from education through economics to environmental sciences. As a consequence, there have been many scientific research papers about coronavirus. As technology and humans’ knowledge to coronavirus advancing, the research output volume is rising over time. Therefore, a bibliographic analysis about COVID-19 related scientific papers will be useful to have an overview of the research progress. \\\\
The main purpose of this analysis is to identify the research trend, collaboration pattern, most influential elements, etc. from publications related to COVID-19 in 2020. We aim to answer the following questions:
\begin{enumerate}
    \item What are the most popular research areas? How has the quantitative output changed over time?
    \item What are the keywords that describe the topics for each period? Is there any switch of research focus?
        \item In the co-authorship network, who are the most frequently collaborated co-authors?
    \item How does the ratio of multidisciplinary paper change over time? Is there any relation to the collaboration pattern?
    \item What are the most cited papers? And what are the corresponding research areas?
    \item Which author has the most significant impact?\\
\end{enumerate}
In Chapter 2, literature reviews of a few previous similar works will be shown. In Chapter 3, methods used in bibliographic analysis and network analysis will be introduced. In Chapter 4, the data refining steps for the analysis will be introduced. In Chapter 5, the results and the analysis will be shown. And finally, in Chapter 6, conclusions from this analysis will be presented.
\section{Literature review}
In this section, a few of recent publications published in 2020 of bibliographic analysis on coronavirus will be reviewed by examining and comparing the methods of data collection, techniques used for visualizations and key findings [2-19].\\
\\
Scientometrics method, which is a sub-field of bibliometric methods, is used to analyse the output. Major data sources for scientific publications are Scopus, Web of Science (WoS), Pubmed, Thomas Reuters, etc. Some papers are built on a single source \cite{safety} \cite{50glob}, other papers use multiple sources. In the latter case, data column discrepancy should be considered and merged accordingly. A fraction of researchers has considered gathering only the documents published in English \cite{bibmapping}, since it is the most prevalent language used in the scientific community. However, most researchers choose to include documents in all languages \cite{iran50}. Depending on one’s own objective, time span for data used in each study is different. Many bibliographic analysis papers focused on scientific publications about COVID-19 selected the data from the beginning of 2020 until recent days before the date of publication in 2020, which unfortunately, results in small amount of data (usually less than 1 thousand), which may lead to bias to the analysis output \cite{iran923}. On the other hand, there are papers focusing on scientific publications about COVID-19 that also include prior 1- or 2-years amount of data \cite{bibglob}, which will contribute little, since the COVID-19 outbreak started barely at the end of December 2019, and the first publication was published in January 2020. Publications related to COVID-19 were gradually published ever since. However, there are examples of large-scale analysis about the coronavirus family as a whole \cite{50glob}\cite{coromapping}\cite{50bibasses}\cite{scientrend}, although the objective is rather different in comparison with this paper’s, the method is worthwhile studying. Almost all of the large-scale studies used data from as early as possible until 2019 or recent days before publication, and usually the amount of data ranges from 10,000 to as many as 45 million. \\
\\
Another concern regarding the preciseness of the analysis result is the data cleaning process. While there are authors who mentioned how they dealt with and aggregated the columns of the data, especially in case where regional geographical data are considered \cite{coromapping}.\\
\\
Many researchers did not disclose the tool for conducting descriptive statistical analysis, but some did release the code [8]. Some common mentioned tools are Excel [4][8][10], borrowing from other researchers [6], or MS-office, Word Cloud generator and ArcGIS software \cite{iran50}. Given the limitations of Excel, in this paper, Python 3 programming language is used. VOSviewer is used among most of the researchers. VOSviewer is used to analyse the relations among highly cited references and productive authors. It is commonly used for mapping and clustering of co-citation network analysis. It also clusters citation terms and portrays the key words by colour [8], etc. Given its convenience, fast speed, and full functionalities, it is indeed an ideal tool for visualizing scientific mappings.\\
\\
Given the nature of bibliographic analysis, the result of one analysis heavily depends on the amount of data collected, especially the time span the data have, as well as the data source chosen. In general, there is no universal conclusion, but very common findings include, the number of scientific publications related to coronaviruses rises sharply after each outbreak in history, and decreases shortly after the peak \cite{scientrend}.\\
\\
Compared to previous works mentioned above, the uniqueness of this paper is that it uses bibliographic data for only 2020. In combination with relatively large amount of data, more accurate results can be achieved, especially the topic is more focused, in this case, only topics related to COVID-19. And moreover, it is also a combination of bibliographic analysis and network analysis, which is relatively more thoroughly. 
\section{Method}
Here we collect the most important concepts in bibliographic analysis and network analysis that we will use throughout the paper.
\subsection{Bibliographic analysis}
\subsubsection{h-index}
The author-level metric h-index is defined as the maximum value of $h$ such that the given author or journal has published $h$ papers that have each been cited at least $h$ times.
$$ \textrm{h-index}( f ) = \max \{ i \in N : f( i ) \geq i \} $$ 
where f is the function that corresponds to the number of citations for each publication ordered in decreasing order \cite{hindexdef}.\\\\
The index is designed to improve upon simpler measures such as the total number of citations or publications. The index works best when comparing scholars working in the same field, since citation conventions differ widely among different fields \cite{hindex}.
\subsubsection{g-index}
The author-level metric g-index $g$ is defined as the largest number $N$ of highly cited articles for which the average number of citations is at least $N$.
$$g^2 \leq \displaystyle\sum_{i \leq g}c_i$$ as 
$$g \leq \frac{1}{g}\displaystyle\sum_{i \leq g}c_i$$
where $c_i$ is number of citations at index $i$ \cite{wikigindex}.
\\
\\
The g-index is an alternative for the older h-index, which does not average the numbers of citations. The h-index only requires a minimum of $N$ citations for the least-cited article in the set and thus ignores the citation count of very highly cited papers. Roughly, the effect is that $h$ is the number of papers of a quality threshold that rises as $h$ rises; $g$ allows citations from higher-cited papers to be used to bolster lower-cited papers in meeting this threshold. Therefore, in all cases $g$ is at least $h$, and is in most cases higher \cite{gindex}.
\subsubsection{Pearson's correlation coefficient (for a population)}
Pearson's correlation coefficient, when applied to a population, is commonly represented by $\rho$:
$$\rho = \frac{cov(X,Y)}{\sigma_X \sigma_Y} $$
where cov is the covariance, $\sigma_X$ is the standard deviation of X, $\sigma_Y$ is the standard deviation of Y.
\subsection{Network analysis}
Network analysis is a set of integrated techniques to depict relations among actors and to analyze the social structures that emerge from the recurrence of these relations \cite{na}. An application of network analysis in social relations is social network analysis. Social network analysis is the process of investigating social structures through the use of networks and graph theory \cite{sna}. Social network analysis has emerged as a key technique in modern sociology. It has also gained significant popularity in biology \cite{biology}, etc.
\subsubsection{Co-authorship network}
Co-authorship network is a complex network where each node represents one author, and there is a link between two nodes if the corresponding authors collaborated in at least one paper.
\subsubsection{Country collaboration network} 
Country collaboration network is a complex network where each node represents one country, and there is a link between two nodes if the two institutions' corresponding countries collaborated in at least one paper.
    \subsubsection{Institution collaboration network}  
Institution collaboration network is a complex network where each node represents an institution, and there is a link between two nodes if two institutions collaborated in at least one paper.
\subsubsection{Research area co-occurrence network}
Research area co-occurrence network is a complex network where each node represents one research area, and there is a link between two nodes if two research areas co-occurrent in at least one paper.
\subsubsection{Keyword co-occurrence network}
Keyword co-occurrence network is a complex network where each node represents one keyword, and there is a link between two nodes if two keywords appear in at least one paper.
\subsubsection{Centrality}
Centrality indicators identify the most important node within the network. Centrality concepts were first developed in social network analysis, and many of the terms used to measure centralities that reflect their sociological origin \cite{socio}. A few centrality measures used in this paper are the following:
\paragraph{Degree centrality}
    Degree centrality calculates an importance score based on the number of links possessed by each node. It is a simple count of the total number of connections linked to a vertex \cite{degcen}.
\paragraph{Betweenness centrality}
    In a graph $G$, the betweenness centrality of a node $v \in V$ is given by the expression:\\
$$g(v) =\displaystyle\sum_{s \neq v \neq t}\frac{\sigma_{st}(v)}{\sigma_{st}},\textrm{ } g \in [0,1] $$ \\
where $\sigma_{st}$ is the total number of shortest paths from node $s$ to node $t$ and $\sigma_{st}(v)$is the number of those paths that pass through $v$. \\
\\    Betweenness centrality measures the number of times a node lies on the shortest path between other nodes. The shortest path is the path with the least number of edges. It is a method of finding 'bridges'. First, it identifies all the shortest paths in the network, and then count how many times each node lies on one and assign a score to it. 
\paragraph{Closeness centrality}
In a connected graph, closeness centrality is calculated as the reciprocal of the farness by Bavelas (1950) times the number of the nodes $N$ in the graph \cite{close1} \cite{close2}.
$$C(x) = \frac{N}{\sum_y d(y,x)}$$ where $d(y,x)$ is the distance of nodes $y$ and $x$. In general, the more central a node is, the closer it is to all other nodes.
\\\\
If the sum of the distances is large, then the closeness is small and vice versa. A vertex with a high closeness centrality would mean it has close relationships with many vertices \cite{graphth}.
\subsubsection{Small-world network}
A small-world network is defined to be a network where the typical distance or shortest length path L between two randomly chosen nodes (the number of steps required) grows proportionally to the logarithm of the number of nodes N in the network, that is:
$$L \propto \log N$$
where N is the number of nodes.\\
\\
Networks with small-world property are rather ubiquitous, such as networks of brain neurons, airport network \cite{airportnw}, social influence networks \cite{socialinf}, telephone call graphs, etc. One property of small-world networks is that it tends to contain cliques, another one is that most pairs of nodes are connected by at least one short path.
\subsubsection{Scale-free network}
A scale-free network is a network whose degree distribution follows a power law, at least asymptotically. That is, the fraction $P(k)$ of nodes in the network having $k$ connections to other nodes goes for large values of $k$ as
$$P(k) \sim k^{(-\gamma)}$$
where $\gamma$  is a parameter whose value is typically in the range $2 < \gamma  < 3$ \\\\
The power law distribution suggests that it is common for a node to have degree exceeding the average. One property of scale-free network is that the clustering coefficient decreases as the degree of node increases, that is, the nodes having low degrees belong to very dense sub-graphs which are connected by hubs. Examples of scale-free networks including airline networks, collaboration networks, webgraph networks, etc.
\subsubsection{Assortativity}
The assortativity coefficient (or Pearson correlation coefficient) is given by
$$r = \frac{\sum_{jk}jk(e_{jk} \textrm{ } q_j q_k)}{\sigma_q^2}$$ 
$q_k$ is the distribution of the remaining degree, $e_{jk}$ refers to the joint probability distribution of the remaining degrees of the two vertices. This quantity is symmetric on an undirected graph, and follows the sum rules
$$\sum_{jk} e_{jk} = 1 \textrm{ and } \displaystyle\sum_{j} e_{jk} = q_k$$ where
$$q_k = \frac{(k+1)p_{k+1}}{\sum_{j \geq 1}jp_j}$$
\section{Data}
\subsection{Data retrieval}
All the data used for analysis are retrieved from Web of Science (WoS), a comprehensive database that provides data for many different academic disciplines, such as mathematics and physics. In WoS, data from core collections are selected. Further, they are filtered by matching topic keywords: covid OR coronavirus OR covid-19 OR covid19 OR 2019-nCoV OR SARS-CoV-2. Since SARS-CoV-2 is the virus that causes COVID-19, it is worthwhile to include it as well. From them, data with time span ranging from 2020 to 2020, whole-year worth amount of data, are selected. Data filtering step can be seen in Figure~\ref{fig:wos_sc} below. In total, 86,834 records of raw data are downloaded as .txt format. Data retrieving date: 8 April, 2021.\\
\begin{figure}[hbt!]
    \centering
    \includegraphics[width=\linewidth]{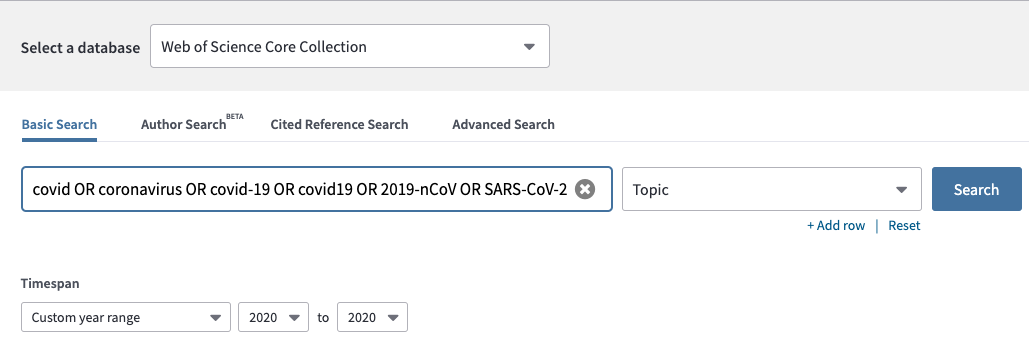}
    \caption{Data filtering}
    \label{fig:wos_sc}
\end{figure}
\subsection{Data selecting}
In Jupyter notebook environment, downloaded data are loaded and fed to a Python function as provided by \cite{20nws} to check duplicates that may come from data retrieving process. 0 duplicates are found. \\
\\
By analysing the details, important columns including Publication Type, Author Full Name, Document Title, ISO Source Abbreviation, Language,  Document Type, Author Keywords, Abstract, Author Address, Cited Reference Count, Total Times Cited Count, Publication Date, Research Areas and Page Count, are kept.\\
\\
Files are then merged to a single .xlsx file while only keeping the columns that are mentioned above. A pandas data frame named \textbf{data} is generated from it. The total size is 82,892 $\times$ 14.
\subsection{Data cleaning}
In Jupyter notebook environment, a second pandas data frame is generated as following:\\
\\
In Publication Date column, discrepancies are found and they are transformed from pandas object data type to datetime data type: 'year-month'.\\
For example:
  \begin{itemize}
  \item 'SEP 10', 'Sep', 'SEPTEMBER 10', 'September', 'SEPT', and 'SEP.' are all corrected to '2020-9'.
  \item Dates that do not have a specific month such as 'SEP-DEC', are transformed to the first appeared month, in this case, '2020-9'.
  \item Several indistinguishable dates such as 'FAL', 'WIN', 'SUM', are dropped.
\end{itemize}
After converting and dropping null values, the remaining data frame is of size 69,192 $\times$ 14. Since there is a significant data loss, it is saved as a second pandas data frame \textbf{data\_w\_date}, and it is used only when date time is necessary for the analysis.
\subsection{Data exploration and transformation}
This section will show more details about the information contained in some columns and how useful data points are extracted for further analysis.
\subsubsection*{Publication Type}
This field contains 4 categories:
\begin{enumerate}
    \item B = Book
    \item J = Journal
    \item P = Patent
    \item S = Book in Series
\end{enumerate}
\subsubsection*{Document Type}
Major types of documents such as Article, Book Review, Correction, Editorial Material, Letter, Meeting abstract, News Item, Retraction, Review, etc. are found. Although there are a number of them that are of Early Access or Proceeding Paper form, only the nature of the publication type will be  considered. For example, 'Article; Early Access' will be considered as 'Article'.
\subsubsection*{Author Address}
The Author Address column contains rather large and complicated information. In each cell, data is represented as the following:\\
\\
\textrm{[Author list]}, Institution, Department, ..., City, Country\\
\\
Institution is a broad term, it can represent a university, a hospital, a research center, etc. that the authors in the list belong to. Followed by the subdivision(s) of the corresponding institution. Then followed by city and country. After examining many rows, a pattern is concluded, in other word, this field gives a department-level information, that is, it is grouped by department, with institution and country information appended accordingly. Hence, one institution can appear multiple times in one cell, so can a country. The goal is to extract institution and country.
\begin{enumerate}
    \item Institutions are extracted by using regular expression. The pattern is that a right square bracket together with a comma \textrm{'],'} marks the beginning of the institution, and a comma \textrm{','} marks the end of the institution. Hence, the following regular expression is used:     \begin{figure}[hbt!]
    \centering
    \includegraphics[width=\linewidth]{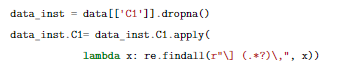}
    \end{figure}
\vspace{5cm}
Two types of institution data are kept:
\begin{itemize}
      \item institution\_data\_unique: only unique institution names appear in one cell will be kept, for analysing quantitative output of each institution.
      \item institution\_data\_nw: keep every institution names from the data, for generating network and performing link analysis.
  \end{itemize}
\item Countries are extracted in a different way, since the form is rather irregular. First, the country subsection is selected by splitting the string. Then, the following transformation applies:
\begin{itemize}
    
  \item The postal code is always prefixed in a US country. For example, 'NJ 08540 USA'. Any string that contains 'USA' will be transformed to 'USA'.
  \item 4 countries: 'North Ireland', 'Wales', 'Scotland', and 'England', are corrected to 'United Kingdom'.
  \item 'P. R. China' and 'Peoples R China' are transformed to 'China'.
  \item 'Viet Nam' and 'Vietnam' are combined together to 'Vietnam'.
  \end{itemize}
  Note that in one cell, one country can appear multiple times, since different institutions can belong to the same country. Similar to institution, two different sets of country data are needed:
  \begin{itemize}
      \item country\_data\_unique: only unique country names appear in one cell will be kept, for analysing quantitative output of each country.
      \item country\_data\_nw: keep every country name from the data, for generating network and performing link analysis.
  \end{itemize}
\end{enumerate}
\subsubsection*{Author Full Names}
 In the Author Full Name column, 874 records of  \textrm{'[anonymous]'} are identified, and they are regarded as null.\\
 \\
However, further issues are found:\\
One author may use the name 'Rodriguez-Jimenez, P.' for one publication, but may use 'Rodriguez-Jimenez, Pedro' for another. Or, sometimes the difference is only a hyphen or a comma. The goal here is to examine whether or not there are duplicate authors. The way of achieving so is by examining string difference between each author pair. If the difference scores above some threshold, then the author pair will be printed for further human examination. Hence, Python library fuzzywuzzy serves as an ideal tool. Fuzzywuzzy calculates Levenshtein distance of two strings and gives scores of different evaluation criteria. Levenshtein distance of string a (length $| a |$) and b (length $| b |$) is given by $lev(a,b)$ where\\
\[
lev(a,b) = 
\begin{cases}
  | a |     &\quad\text{if  } | b | = 0\\
  | b |     &\quad\text{if  } | a | = 0\\
  lev(tail(a), tail(b)) &\quad\text{if  } a[0] = b[0]\\
  1 + \min \begin{cases}
    lev(tail(a), b)\\
    lev(a, tail(b))\\
    lev(tail(a), tail(b))\\
  \end{cases}
  &\quad\text{otherwise}
\end{cases}
\]\\
where $tail(x)$ is the remaining string of $x$ after deleting the first character, and $x[n]$ is the nth character of $x$, starting from character 0.\\
\\
In fact, by setting the Levenshtein distance ratio score's threshold to 0.8, after comparing string similarities of each unique author pair in a rather small fraction of the whole author data, many errors are identified. Due to the huge volume of the data, it is nearly impossible to correct each of them. Moreover, this error is negligent \cite{20nws}. Hence, no author names are corrected.
\section{Result}
In this section, full result of the analysis will be shown. The majority of the analysis is conducted in Jupyter notebook. In addition, VOSviewer is used for generating network visualizations.
\subsection{Descriptive data analysis}
Given the simplicity, all of the descriptive data analysis are conducted in Jupyter notebook. In order to do so,  standard Python libraries: pandas, numpy, matplotlib.pyplot, seaborn, datetime, re, os, are imported.
\subsubsection{Publication type}
Out of 82,892 records, 82,888 of them are Journals, 3 of them are Book in Series, 1 of them is Book.
\subsubsection{Document type}
Having introduced in the previous section, first, in document type column, data are corrected accordingly. Document type can be regarded as categorical data. After taking a detailed look, 16 different categories are identified. As the figure below shows, the majority of the papers are articles, which take part in 51.3\% of the whole publications, followed by editorial materials, 17.2\%, then letter, 16.3\%.
\begin{figure}[h!]
    \centering
    \includegraphics[width=0.9\textwidth,scale=0.2]{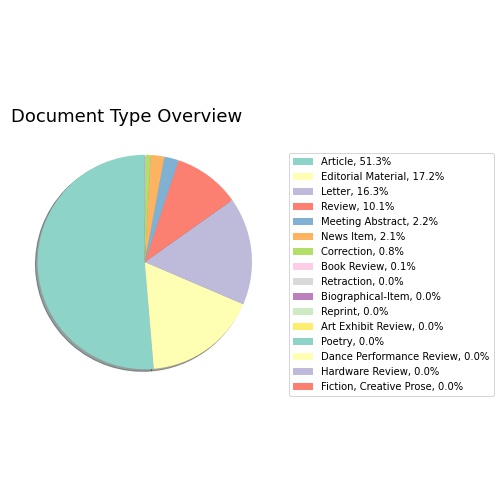}
    \caption{Document type}
    \label{fig:doctype}
\end{figure}
\pagebreak
\clearpage
\subsubsection{Language}
Data for visualizing language are derived from the Language column in the data set. Based on the initial observation, each paper has been written in only one language. Hence, it is of categorical type, each category represents one language. In total, 31 languages are identified. Given the large number of categories, only ten most used languages are selected for visualization. As the figure below shows, English as an international language, is the mostly used language. 
\begin{figure}[h!]
    \centering
    \includegraphics[width=\linewidth]{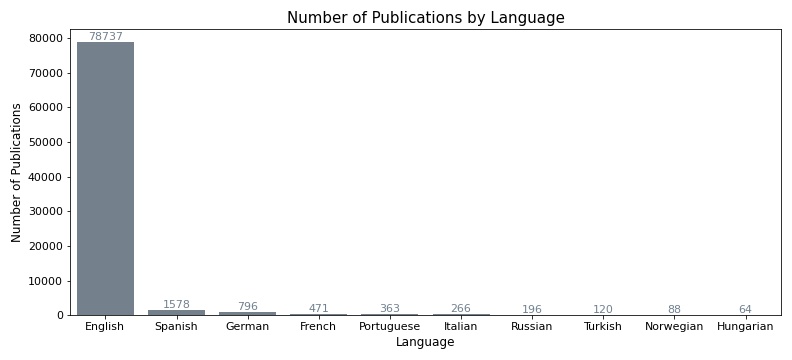}
    \caption{Language}
    \label{fig:doclang}
\end{figure}
\subsubsection{Publication source}
Publication source data are derived from ISO Source Abbreviation column from the data set. Moreover, it is the shorter version of Publication Name column in the initial data set. Since for some publication sources, the name is rather long, which will create difficulties in the later visualization step, the equivalent abbreviation form is used. It can be regarded as categorical data. In total, 7,501 publication sources are identified. Given the large volume, they are further being sorted in decreasing order by publication quantity, only 10 publication sources that published the most are selected. Bar plot is used for visualization. A list that contains the above ten publication sources is also created for further analysis related to time change. As the figure shows, BMJ-British Medical Journal is the leading source.
\begin{figure}[h!]
    \centering
    \includegraphics[width=\linewidth]{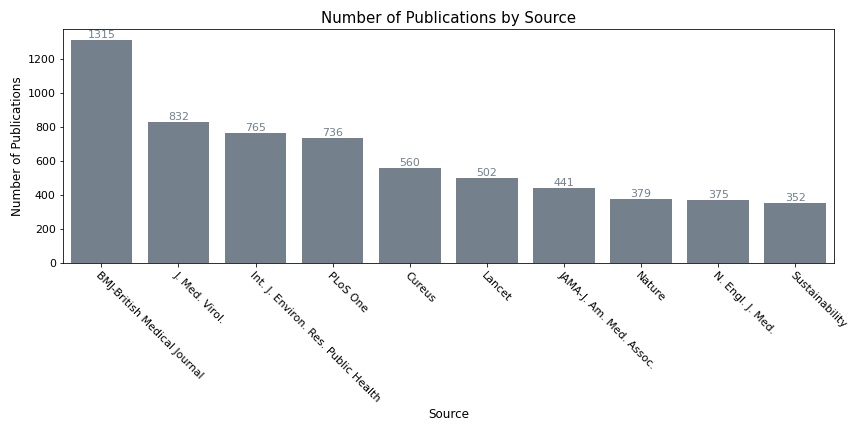}
    \caption{Publication source}
    \label{fig:docsource}
\end{figure}
\pagebreak
\subsubsection{Number of pages}
The data being used are from the column Page Count. It is clearly of numerical type. The least number of pages is 1, the 5 largest number of pages are: 423, 310, 293, 172, 161. On average, one paper has 6.99 pages, and the median number is 5. The most common page number is 2.
\begin{figure}[h!]
    \centering
    \includegraphics[width=0.5\linewidth]{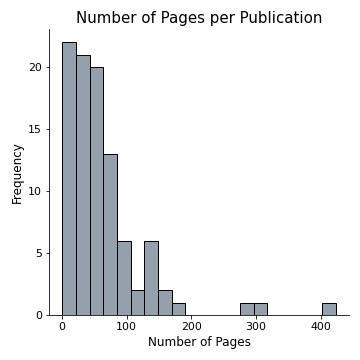}
    \caption{Page count}
    \label{fig:docpage}
\end{figure}
\pagebreak
\subsubsection{Country}
Country data country\_data\_unique are derived and cleaned from Author Address column as being introduced in the previous section. In total, there are 201 countries identified in the data set, however, only the top ten countries from the ranking list by number of publications will be selected, since this will tell what the most productive countries are. A list that contains the top 10 productive countries is created for further analysis over month. As the figure shows, USA as a country with strong academic research, has the most publications, followed by China, UK, and Italy.
\begin{figure}[h!]
    \centering
    \includegraphics[width=\linewidth]{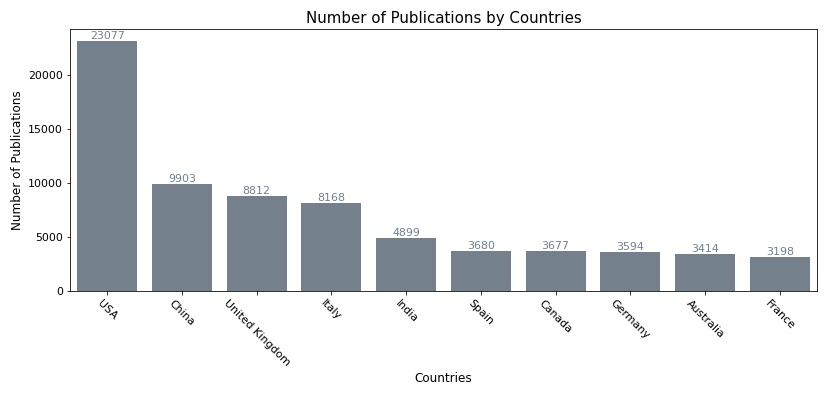}
    \caption{Country}
    \label{fig:doccountry}
\end{figure}
\pagebreak
\subsubsection{Institution}
Similarly, data institution\_data\_unique are also derived and cleaned from Author Address column. Repeating the same steps, 52,421 institutions are identified, only top 10 institutions from the ranking list will be selected. As the figure shows below, with no surprise, many top institutions can be seen here. Harvard Medical School in USA is the most productive institution, followed by Huazhong University of Science \& Technology, which is a top Chinese research university located in Wuhan. The third institution in the ranking list is a Canadian university.
\begin{figure}[h!]
    \centering
    \includegraphics[width=\linewidth]{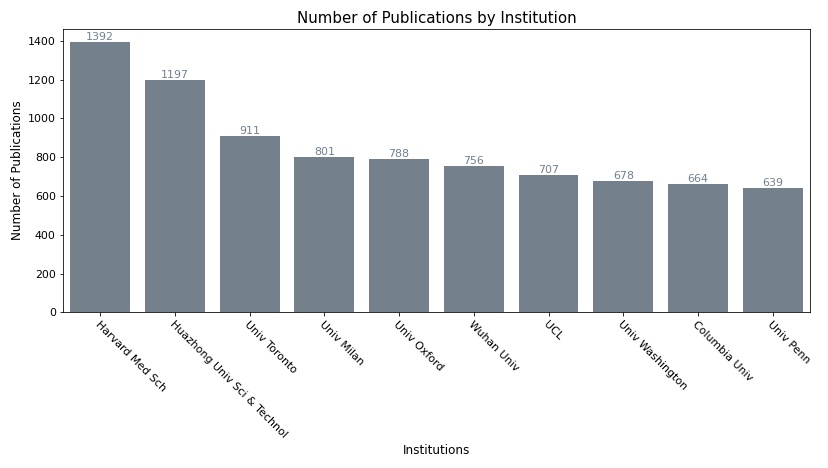}
    \caption{Institution}
    \label{fig:docinst}
\end{figure}
\pagebreak
\subsubsection{Research area}
The data are derived from Research Areas column. In total, 151 different research areas are identified. Only the ten research areas that appears the most will be selected. From the figure, it can be seen that General \& Internal Medicine is the most sought after research area. Since the main purpose of the research is to battle the virus. It is also surprising to see that psychiatry is also among the most popular research areas.
\begin{figure}[h!]
    \centering
    \includegraphics[width=\linewidth]{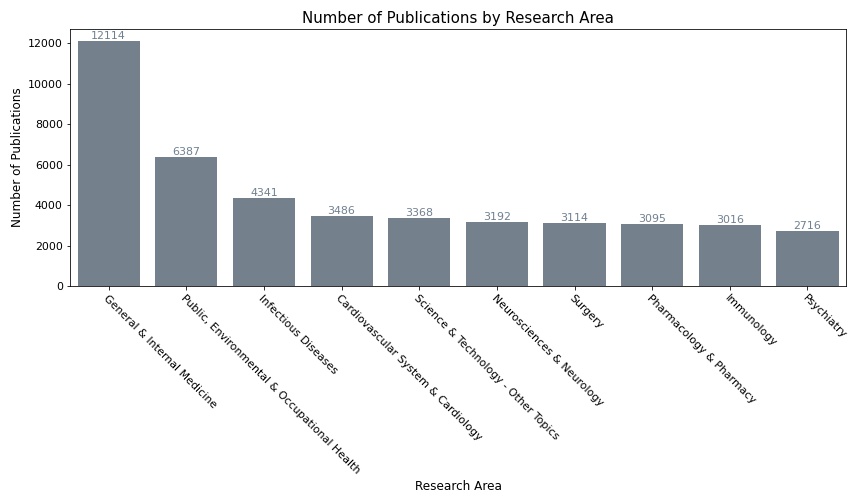}
    \caption{Research area}
    \label{fig:docresearch}
\end{figure}
\pagebreak
\subsubsection{Author keyword}
The data are derived from Author Keywords column. Author keywords highlight one author's research focus, hence, it is useful to see that among authors, what topics are discussed most frequently, similar to research area, however, keywords will give a more divided category. In total, 62,565 author keywords are identified. Among them, 20 most frequently appeared keywords will be selected. In addition to those familiar medical terms, mental health, anxiety and depression also appear among the most frequent keywords. This could be the result of the lockdown that cuts off social interactions, which also affects human's mentality.
\begin{table}[h!]
\begin{tabular}{ |p{4cm}|p{2cm}|p{4cm}|p{2cm}| } 
\hline
\textbf{Author Keywords} &	\textbf{Count} & \textbf{Author Keywords} &	\textbf{Count}\\
\hline
covid-19 & 29,920 &pneumonia	&800\\
\hline
sars-cov-2& 9,442 & anxiety&	777\\
\hline
coronavirus & 6,327 & mortality&	757\\
\hline
pandemic & 4,106 & pandemics&	703\\
\hline
public health	& 1,070 & lockdown&	630\\
\hline
coronavirus disease 2019	& 988 & depression&	575\\
\hline
mental health & 959 & ace2&	560\\
\hline
epidemiology & 828& severe acute respiratory syndrome coronavirus 2	&553\\
\hline
covid-19 pandemic & 818 & hydroxychloroquine&	491\\
\hline
telemedicine & 806& children&	478\\
\hline
\end{tabular}
\caption{Author Keywords}
\label{table:authorkey}
\end{table}
\pagebreak
\subsubsection{Number of authors per paper}
The data are derived and cleaned from Author Full Name column as introduced in previous chapter. From this data, it shows that the least number of authors in one paper is 1, and the largest number is 705, which can be considered as an outlier. On average, 5.75 authors collaborated in one paper, and the median number is 4. The most frequent number of authors in one paper is 1.
\begin{figure}[h!]
    \centering
    \includegraphics[width=0.5\linewidth]{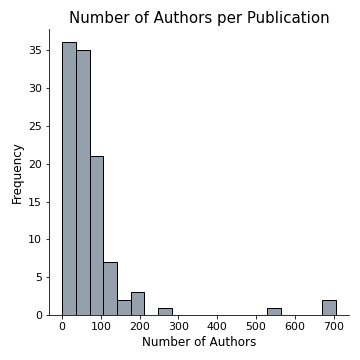}
    \caption{Number of authors per paper}
    \label{fig:auppr}
\end{figure}
\pagebreak

\subsubsection{Most cited papers}
The data are derived from Document Title, Research Areas and Total Times Cited Count columns. Here, on a document-level basis, we want to identify the most popular papers by how many times it has been cited by others. By rearranging the selected data frame in number of citation decreasing order, 10 most cited papers and their corresponding research areas are selected and a table format is generated.

\begin{table}[h!]
\begin{tabular}{|p{0.7\linewidth} | p{0.18\linewidth}|p{0.12\linewidth}|}
\hline

    \textbf{Title} & \textbf{Authors} & \textbf{Number cited}\\
\hline
Clinical features of patients   infected with 2019 novel coronavirus in Wuhan, China& Huang, Chaolin et al. & 10,528\\
\hline
Clinical Characteristics of Coronavirus Disease 2019 in China & Guan, W. et al. & 6,486\\
\hline
Clinical Characteristics of 138 Hospitalized Patients With 2019 Novel Coronavirus-Infected Pneumonia in Wuhan, China & Wang, Dawei et al. &  5,925\\
\hline
A Novel Coronavirus from Patients with Pneumonia in China, 2019& Zhu, Na et al. & 5,596 \\
\hline
Clinical course and risk factors for mortality of adult inpatients with COVID-19 in Wuhan, China: a retrospective cohort study& Zhou, Fei et al. &  5,556\\
\hline
\end{tabular}
\caption{Most cited papers}
\label{table:citedpprs}
\end{table}
\begin{table}[h!]
\begin{tabular}{|p{0.7\linewidth} | p{0.18\linewidth}|p{0.12\linewidth}|}
\hline
\textbf{Title} &\textbf{Authors}&	\textbf{Number cited}\\
\hline
Epidemiological and clinical characteristics of 99 cases of 2019 novel coronavirus pneumonia in Wuhan, China: a descriptive study& Chen, Nanshan et al. &5,354\\
\hline
A pneumonia outbreak associated with a new coronavirus of probable bat origin& Zhou, Peng et al. & 4,365\\
\hline
Early Transmission Dynamics in Wuhan, China, of Novel Coronavirus-Infected Pneumonia& Li, Qun et al. & 3,530\\
\hline
Characteristics of and Important Lessons From the Coronavirus Disease 2019 (COVID-19) Outbreak in China Summary of a Report of 72 314 Cases From the Chinese Center for Disease Control and Prevention& Wu, Zunyou et al. & 3,171\\
\hline
SARS-CoV-2 Cell Entry Depends on ACE2 and TMPRSS2 and Is Blocked by a Clinically Proven Protease Inhibitor& Hoffmann, Markus et al. &  3,030 \\
\hline
\end{tabular}
\caption{Most cited papers (2)}
\label{table:citedpprs2}
\end{table}
\clearpage
\pagebreak
\subsubsection{Author table}
On a author-level basis, a ranking table of measuring author's quantitative impact is produced by calculating the following:
\begin{itemize}
    \item Total number of times cited
    \item Number of papers
    \item Number of times cited per paper
\end{itemize}
By sorting the above three metrics, less bias will be produced, since there are authors who have rather large number of publications, but only being cited a few times, on the other hand, there are also authors who only write few but highly cited papers. From the ranking list, top ten authors are selected. The following table is ranked by the first column.
\begin{table}[h!]
\centering
\begin{tabular}{ |p{4cm}|p{3cm}|p{2.2cm}|p{2.2cm}| } 
\hline
\textbf{Author name} &	\textbf{Total number cited} & \textbf{Cited per paper} &	\textbf{Number of papers}\\
\hline
Yu, Ting&	24,659&	2,054.9&	12\\
\hline
	Wei, Yuan&	21,625	&2,402.8&	9\\
	\hline
	Zhang, Li&	18,410&	354.0&	52\\
	\hline
	Li, Hui&	18,383&	417.8&	44\\
	\hline
	Cao, Bin&	17,799&	773.9&	23\\
	\hline
	Cheng, Zhenshun&	17,285&	1,329.6&	13\\
	\hline
	Xu, Jiuyang	&17,015	&1,546.8	&11\\
	\hline
	Fan, Guohui	&16,804&	1,680.4	&10\\
	\hline
	Wang, Yeming&	16,793&	1,526.6	&11\\
	\hline
	Gu, Xiaoying&	16,786&	1,865.1	&9\\
	\hline
\end{tabular}
\caption{Author table}
\label{table:authortb}
\end{table}
\pagebreak
\subsection{Data analysis with time}
Unlike Descriptive data analysis, analysing data with time series will normally result in analysing in higher dimensions. In contrast to 1D analysis, such as counting, grouping, summing based on one column, time series analysis often uses data that have more complex structures. It aims at showing dynamic changes of the data across time. It helps to find the pattern, trend, further for predicting, making better decisions. \\\\
In this section, some of the data introduced before are combined with time, to show a basic quantitative overview of the change in this one-year time span, with one month chosen for one unit. The analysis is conducted in Jupyter notebook.
\subsubsection{By month}
The column of transformed publication date is used.
The total quantity of research output over months will be analysed. From the figure below, a progressive growing trend can be seen. From February to July, the number of publications grows gradually. From July, the growing speed slows down, a declining trend can also be observed, but the number of publications in each month still remains in high volume. 
\begin{figure}[h!]
    \centering
    \includegraphics[width=\linewidth]{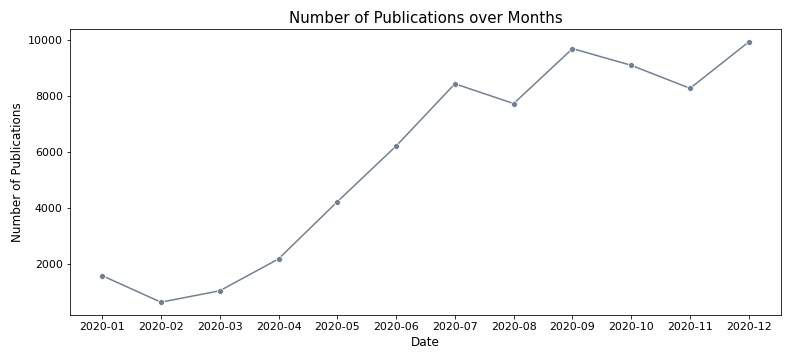}
    \caption{Publications over month}
    \label{fig:datecount}
\end{figure}
\pagebreak
\subsubsection{By country over month}
Columns of Author addresses and Publication date in the data set are used. As mentioned in the previous chapter, the country\_data\_unique data set is used. For the purpose of visualizing, only 10 most productive countries as shown before will be selected. As the figure below shows, the publication trend in USA leads the publication trend worldwide. In other most productive countries, it showed a very slow growing trend in the first half of the year, then stagnated yet still remained in rather high productivity level.
\begin{figure}[h!]
    \centering
    \includegraphics[width=\linewidth]{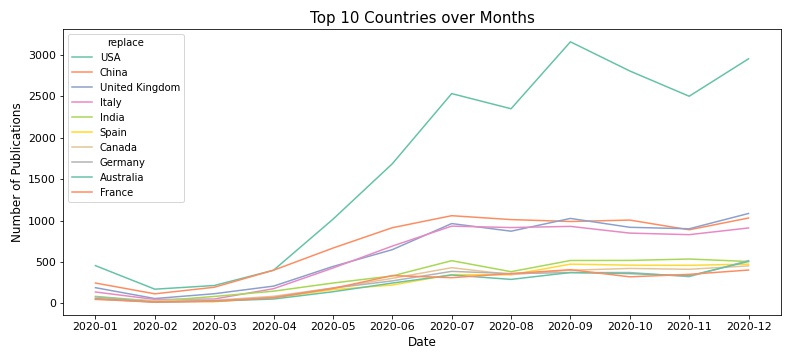}
    \caption{Countries over month}
    \label{fig:country_month_10}
\end{figure}
\pagebreak
\subsubsection{By publication sources over month}
Columns of ISO Abbreviation and Publication date are selected. Only ten most productive publication sources will be selected for visualization. As the figure below suggests, there is no fixed pattern in terms of how one publication source updates its database, even so, a general growing trend can still be seen among many publication sources.
\begin{figure}[h!]
    \centering
    \includegraphics[width=\linewidth]{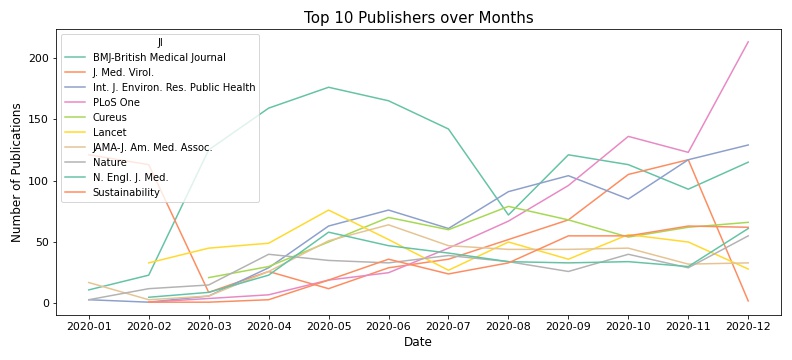}
    \caption{Publication source over month}
    \label{fig:publisher_month}
\end{figure}
\pagebreak
\subsubsection{By research area over month}
Columns of Research Area and Publication Date are selected. Only ten most popular research areas will be selected for visualization. As the figure below shows, a clear growing trend can be observed in each most popular research area. Like in the general trend seen before, the first half of 2020 is still the period where the number of publications in each research area significantly increases each month. At the end of 2020, from the figure, it suggests that at the beginning of the following year, the number of publications in those popular research areas will have a significant increase as well.
\begin{figure}[h!]
    \centering
    \includegraphics[width=\linewidth]{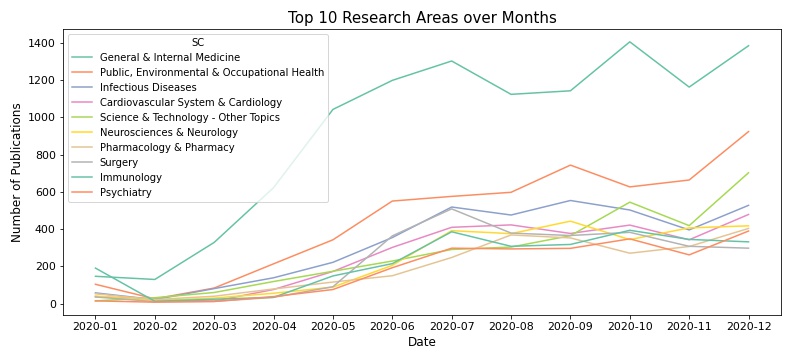}
    \caption{Research area over month}
    \label{fig:research_month}
\end{figure}
\pagebreak
\subsection{Author-level metric}
\subsubsection*{h-index, g-index}
Both h-index and g-index are similar metrics to measure both productivity and citation impact of one author. Compared to the previous author table that measured only the simple quantitative output, this gives a completely different picture. And it is surprising that none of the authors in the previous table appears in this author index table. As the table shown below, the author with the most impact is an Italian author, Lippi Giuseppe.
\begin{table}[h!]
\centering
\begin{tabular}{ |p{6cm}|p{2cm}|p{2cm}| } 
\hline
\textbf{Author name} &	\textbf{h-index} & \textbf{g-index}\\
\hline
Lippi, Giuseppe	&26&	62\\
\hline
	Liu, Lei&	26&	50\\
	\hline
	Drosten, Christian&	23&	38\\
	\hline
	Baric, Ralph S.	&21	&47\\
	\hline
	Baric, Ralph S.Wang, Wei&	20	&56\\
	\hline
	Zhang, Wei	&19	&49\\
	\hline
	Yuen, Kwok-Yung	&18	&46\\
	\hline
	Yang, Yang	&18	&37\\
	\hline
	Rodriguez-Morales, Alfonso J.&	18&	36\\
	\hline
	Wang, Ying	&17	&51\\
	\hline
\end{tabular}
\caption{Author index}
\label{table:authorindex}
\end{table}
\pagebreak
\subsection{Correlation matrix}
A correlation matrix is a table that shows correlation coefficients among the variables. It aims to summarize the large amount of data to characterize relations, such as whether or not two variables are highly correlated with each other. Metrics of correlation coefficients, in this case, is calculated among each pair of variables. The correlation coefficient r ranges from -1 to +1. +1 indicates there is perfect positive relationship between two variables, -1 indicates perfect negative relationship. 0 means no relationship exists. On a document-level basis, as many as the numerical information is extracted, the goal is to produce a correlation matrix to examine if there exhibits relations among them. \\\\
Numerical information being extracted are: Number of authors, cited reference count, total times cited, number of research area, number of countries, and page count. All rows that contain nan values are dropped, so that the remaining data set has only valid value. Pearson correlation coefficient is calculated. \\\\
From the figure below, it can be seen that variables Page Count and Cited Reference Count have the largest correlation coefficient, which suggests that the more pages one publication has, the more papers it cites. The second largest is the correlation between Number of Countries and Number of Authors, which is also very intuitive, since the larger the number of author, the higher the probability that they come from different countries. 
\begin{figure}[h!]
    \centering
    \includegraphics[width=0.8\linewidth]{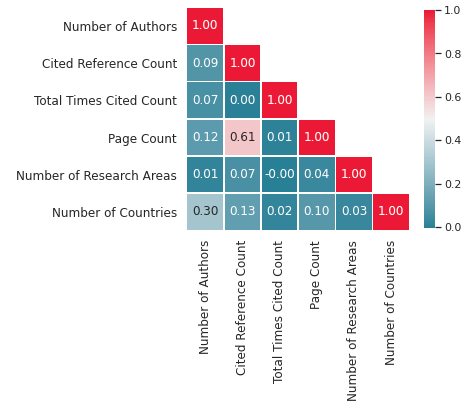}
    \caption{Correlation matrix}
    \label{fig:corr}
\end{figure}
\pagebreak
\subsection{Degree of collaboration}
The degree of collaboration of a given period is defined as\\
$$\frac{\{\sum \textrm{papers with \# authors} \geq 1\}}{\{\sum \textrm{papers with \# authors} \geq 1\}+\{\sum \textrm{papers with \# authors} = 1\}}$$\\
A paper is called a collaborated paper if it is written by more than one author, and a sole paper if it's written by only one author. Although as the previous analysis showed, the number of papers written by only one author takes only a small fraction, it is still useful for analysing collaboration trend. From the figure below, it can be seen that throughout the year, the degree of collaboration remained high, at above 0.5. Although there was a downwards trend around March, the reason is unclear. One of the possibilities is that it was when the lockdown in Europe has started, which increased the isolation. After that, the degree of collaboration rose gradually, to the end of the year, it reached to a plateau, and a small downwards trend can be observed.
\begin{figure}[h!]
    \centering
    \includegraphics[width=\linewidth]{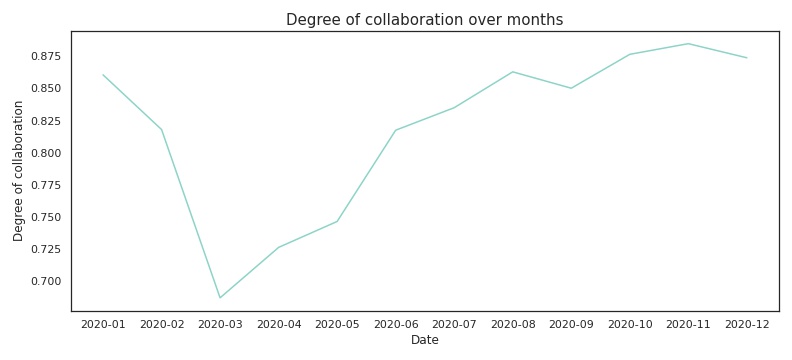}
    \caption{Degree of collaboration}
    \label{fig:degcolab}
\end{figure}
\pagebreak
\subsection{International collaboration and Multidisciplinary}
A paper is called an internationally collaborated paper if there are more than one country where the corresponding institution belongs collaborated together. Similarly, a paper is called a multidisciplinary paper if there are more than one research area that occur in it. From the figure below, it can be seen that throughout the year, the international collaboration remained at the same level, at around 0.225. It also tells that domestic collaboration remained prevalent. However the multidisciplinary ratio is also low throughout the year, and it is particularly low around from February to April. One of the explanations could be, that after the virus has spread worldwide, researchers and scientists started to concentrate on the fundamental biology of the virus, in order to come up with solutions to the pandemic. But in general, the multidisciplinary trend cannot be concluded.
\begin{figure}[h!]
    \centering
    \includegraphics[width=\linewidth]{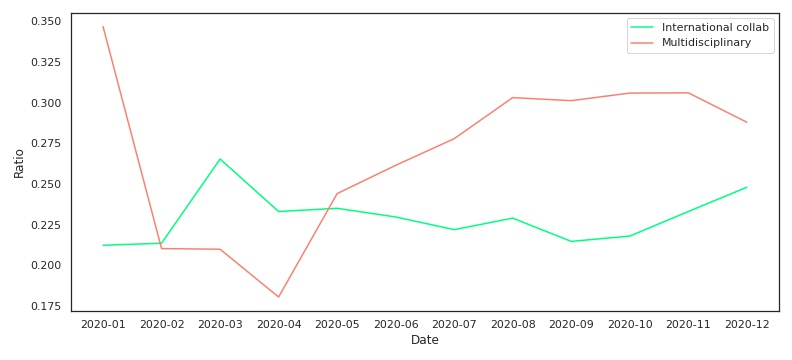}
    \caption{International collaboration and Multidisciplinary}
    \label{fig:int_colab_ratio}
\end{figure}
\pagebreak
\subsection{Keyword visualization}
The goal of keyword visualization is to extract most common keywords and generate a worldcloud containing them. The 'Author Keywords' column in the original data set is provided by WoS. Although it has its own advantage of having a more systematic keyword dictionary, the limited amount of keywords will not be ideal for providing more detailed topics. Hence, another approach is needed to support enough amount of topic words.
\subsubsection{NLTK}
In this section, Natural Language Toolkit (NLTK) is used to extract keywords from 'Document title' and 'Abstract' columns, which will provide the fundamental data for generating word cloud.\\\\
The Natural Language Toolkit, or more commonly NLTK, is a suite of libraries and programs for natural language processing (NLP). It is essential to combine the title and the abstract columns, so that for each publication, there is a single list of strings that represents its topics. \\\\
Several steps are needed when processing the text data in order to find the keywords:
\begin{itemize}
  \item The word appears at the beginning of the sentence is always capitalized, and it will be considered as a different word from the non-capitalized version. Hence, all the letters are transformed to lower.
  \item A set of stop words is created by combining English stop words, punctuation, and digits. Since by using only English stop words, many trivial characters are found among the most frequent keywords.
  \item The keywords are selected by removing stop words created in step two, and the frequency of appearance is calculated for each keyword, then they are sorted in descending order by frequency.
\end{itemize}
\subsubsection{Wordcloud}
Wordcloud is a Python library that generates word clouds from the vocabularies provided based on the frequency they appear for instance. The more frequent a word appears, the bigger the font it displays in the word cloud. It also allows customization by adjusting the amount of vocabulary to keep. The purpose is to give a better visualized picture that shows the most frequent topics related to the text. And it can be used for further advanced structural analysis. \\
\\
Having had the method to fetch the most frequent words, Word clouds can be generated from them. The 100 most frequent keywords in the whole period from the data set will be used.
\begin{figure}[h!]
    \centering
    \includegraphics[width=1.1\linewidth]{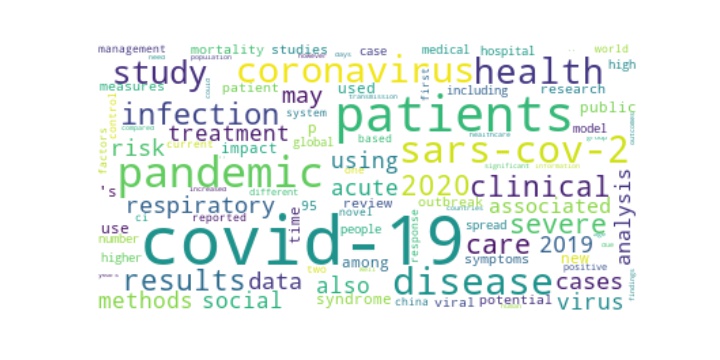}
    \caption{Wordcloud 2020}
    \label{fig:cloud2020}
\end{figure}
\pagebreak
\subsection{Network analysis}
Having introduced in Chapter 3, results of network analysis will be shown here.
\subsubsection{NetworkX}
NetworkX is a Python library to create, manipulate, and study of the complex network's structure, dynamics,functions, etc.
NetworkX provides tools for creating networks from existing data sets,  certain graphs with fixed parameters, or from manual input. It can also generate random graphs. Hence, it is useful for studying graph structures, designing algorithm, building models, drawing graphs, etc.\\\\
For the purpose of this paper, NetworkX is used for generating complex networks for different purposes and the calculations related to the property measures of the network.\\\\
The above NetworkX networks are generated from pandas data frame, nodes, edges are added row by row. This will generate networks consisting of undirected multigraphs, with self-loops, multiple edges allowed, with one attribute being 'weight', which represents the number of edges attached to the node.
\subsubsection{Co-authorship network}
Co-authorship network is a complex network where each node represents an author, two nodes are connected by a link if there is at least one collaboration between two authors. There are possibilities that one author pair has collaborated more than once, hence, weighted edges are formed.\\\\
By applying functions from NetworkX, Some facts are calculated.\\Facts of the co-authorship network:
\begin{itemize}
    \item Number of nodes: 304,880
    \item Number of edges: 2,644,926
    \item Number of isolated nodes: 7,779
    \item Number of connected components: 21,223
    \item Size of the largest component: 189,790
    \item Size of top 10 largest components: 189,790; 128; 123; 103; 102; 93; 92; 90; 87; 73
\end{itemize}
From above, it can be seen that there are 304,880 unique authors contributing to publications related to COVID-19. Out of them, 7,779 have written papers solely on their own and have never participated in collaborated papers. And in total, there are 2,644,926 collaborations among them. The largest component consists of more than 63\% of the total authors. 21,223 number of connected components also suggests that there are rather large number of collaboration communities, and the majority of the authors belongs to the top largest communities.\\
The most collaborative author pairs are shown below, Kow, Chia Siang and Hasan, Syed Shahzad have the most frequent collaborations.
\begin{table}[h!]
\centering
\begin{tabular}{ |p{5cm}|p{5.2cm}|p{2.3cm}| } 
\hline
\textbf{Author name} &	\textbf{Author name} & \textbf{Count}\\
\hline
Kow, Chia Siang	&Hasan, Syed Shahzad&	{'weight': 36}\\
\hline
	Joob, Beuy&	Wiwanitkit, Viroj&	{'weight': 35}\\
	\hline
	Dhama, Kuldeep&	Rodriguez-Morales, Alfonso J.&	{'weight': 33}\\
	\hline
	Yuen, Kwok-Yung&	Chan, Jasper Fuk-Woo&	{'weight': 33}\\
	\hline
	Tiwari, Ruchi&	Dhama, Kuldeep	&{'weight': 32}\\
	\hline
	To, Kelvin Kai-Wang&	Yuen, Kwok-Yung&	{'weight': 31}\\
	\hline
	Lechien, Jerome R.&	Saussez, Sven&	{'weight': 31}\\
	\hline
	Lippi, Giuseppe&	Henry, Brandon Michael&	{'weight': 30}\\
	\hline
	Sah, Ranjit&	Rodriguez-Morales, Alfonso J.	&{'weight': 29}\\
	\hline
	He, Daihai&	Zhao, Shi&	{'weight': 29}\\
	\hline
\end{tabular}
\caption{Co-author pairs}
\label{table:coaup}
\end{table}
\begin{figure}[hbt!]
    \centering
    \includegraphics[width=1.03\linewidth]{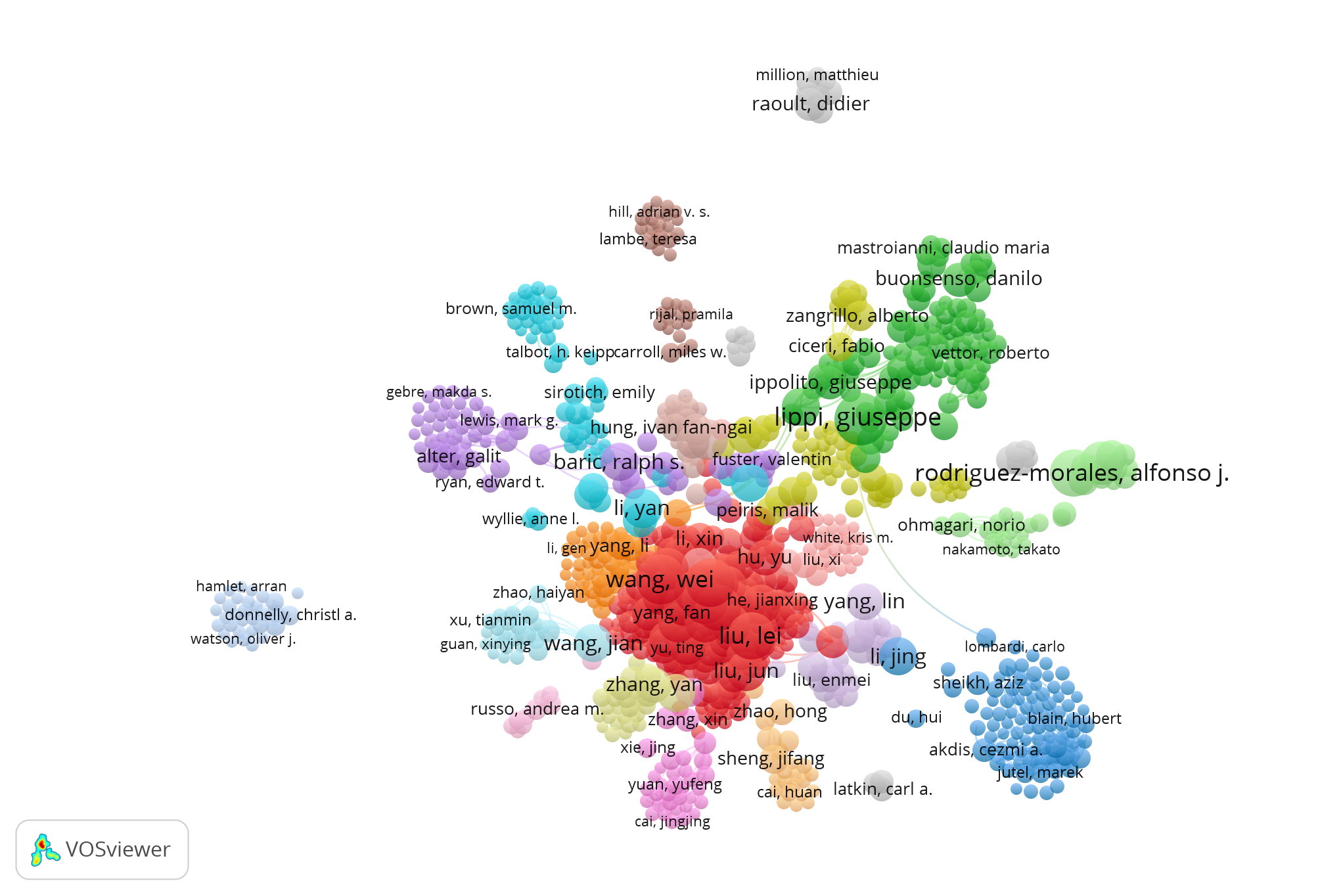}
    \caption{Network Visualization of co-authorship network}
    \label{fig:auvos}
\end{figure}
\clearpage
\subsubsection{Country collaboration network}
Country-level collaboration network is a complex network where each node represents a country, two nodes are connected by a link if there is at least one collaboration between those two institutions' corresponding countries. Similar to the co-authorship network, in this country collaboration network, since two countries can collaborate in at least one paper, weighted edges are formed. 
By applying functions from NetworkX, some facts are calculated.\\
Facts of the Country co-occurrence network:
\begin{itemize}
    \item Number of nodes: 201
    \item Number of edges: 6,404
    \item Number of isolated nodes: 5
    \item Number of connected components: 1
    \item Size of the largest component: 196    
\end{itemize}
As shown above, there are 201 countries participated in the publications, with 6,404 collaborations. There are 5 isolated nodes, which means that for those countries, papers are published by a single department of one institution, and there are no international collaborations in those countries. The most collaborative country pairs are shown in the following table. USA has the most collaborations with other countries. The second most collaborative country is UK, who collaborated with USA, Italy, Australia, China, etc.
\begin{table}[hbt!]
\centering
\begin{tabular}{ |p{2cm}|p{2cm}|p{4cm}| } 
\hline
\textbf{Country} &	\textbf{Country} & \textbf{Count}\\
\hline
USA&	USA&	{'weight': 11,164}\\
\hline
	USA&	UK&	{'weight': 1,620}\\
	\hline
	USA&	China&	{'weight': 1,533}\\
	\hline
	USA&	Italy&	{'weight': 1,187}\\
	\hline
	USA&	Canada&	{'weight': 1,158}\\
	\hline
	UK&	Italy&	{'weight': 927}\\
	\hline
	USA&	Australia&	{'weight': 831}\\
	\hline
	USA&	Germany&	{'weight': 742}\\
	\hline
	UK&	Australia&	{'weight': 684}\\
	\hline
	UK&	China&	{'weight': 673}\\

	\hline
\end{tabular}
\caption{Country collaboration pairs}
\label{table:ccp}
\end{table}
\begin{figure}[hbt!]
    \centering
    \includegraphics[width=1.08\linewidth]{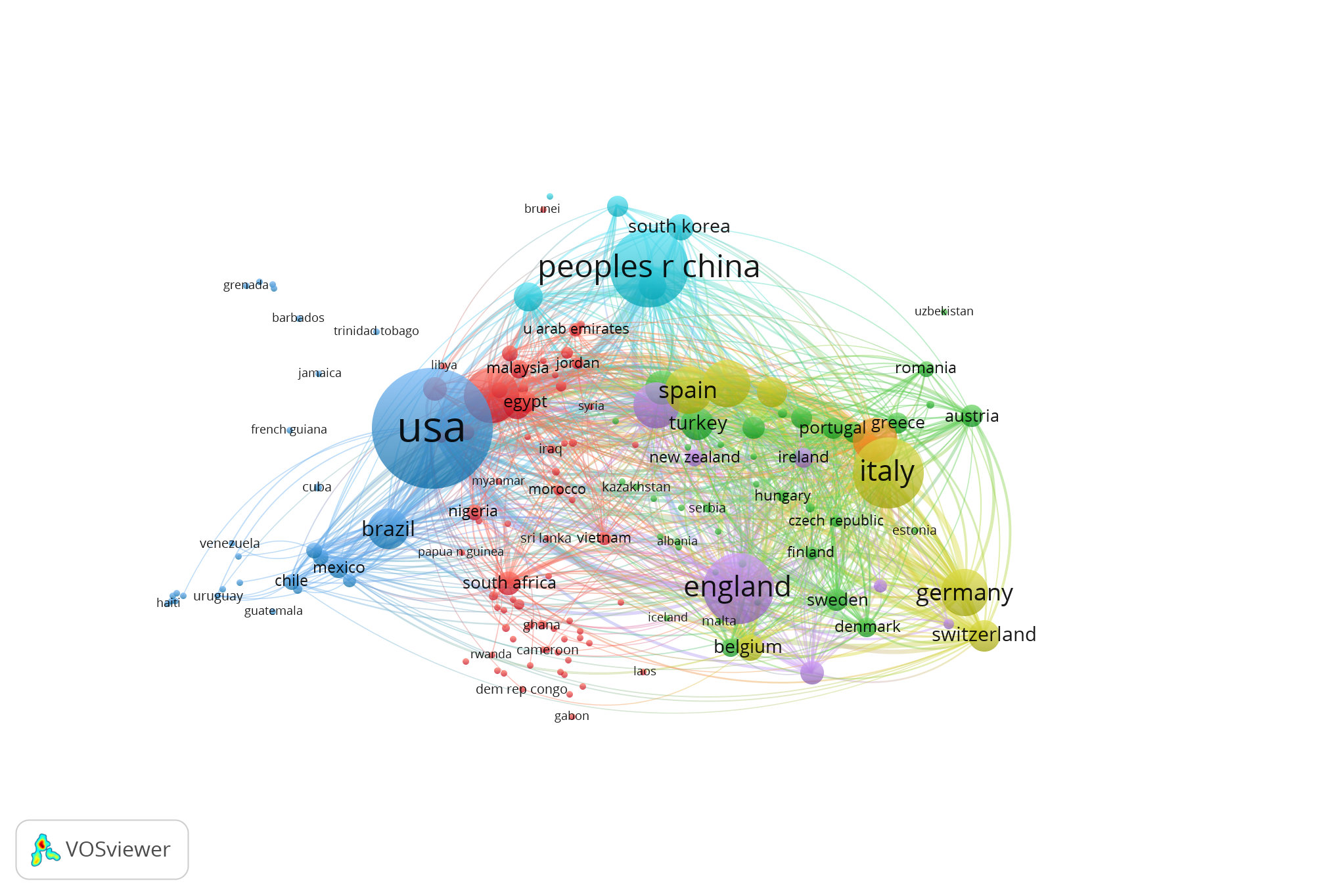}
    \caption{Network Visualization of country collaboration network}
    \label{fig:countryvos}
\end{figure}
\pagebreak
\clearpage
\newpage
\subsubsection{Institution collaboration network}
A institution co-occurrence network is a network where each node represents an institution. There is a link between two nodes if there is one publication collaborated by two institutions. Multiple edges can exist since two institutions can collaborate more than once. Self-loops can exist as well, since one paper can be written by one institution, but from different departments, as the data set suggests. One institution collaborates with itself if more than one of its departments are involved.\\
\\
By applying functions from NetworkX, Some facts are calculated.\\Facts of the Institution co-occurrence network:
\begin{itemize}
    \item Number of nodes: 52,421
    \item Number of edges: 535,398
    \item Number of self-loops: 8,006
    \item Number of isolated nodes: 2,808
    \item Number of connected components: 1,575
    \item Size of the largest component: 46,347
    \item Size of top 10 largest components: 46,347, 19, 18, 14, 13, 11, 9, 9, 8, 8
\end{itemize}
From above, it can be seen that there are 52,421 institutions participated in the publications, and there are 535,398 collaborations in total. Out of them, there are 2808 institutions that do not have any collaborations, either internally, or with other institutions. The most frequent institutional collaborations exists in Huazhong Univ Sci \& Technol as internal collaborations, 559 times of collaborations, which is more than double of the internal collaborations in Harvard Med School, which is 277 times. It is also a fact that internal collaborations in institutions remained prevalent. The following table shows the most frequent inter-institutional collaborations.
\begin{table}[h!]
\centering
\begin{tabular}{ |p{5cm}|p{5cm}|p{3cm}| } 
\hline
\textbf{Institution} &	\textbf{Institution} & \textbf{Count}\\
\hline
Harvard Med Sch&	Massachusetts Gen Hospl&	{'weight': 243}\\
\hline
	Harvard Med Sch	&Brigham \& Womens Hosp	&{'weight': 206}\\
	\hline
	Univ Chinese Acad Sci		&Chinese Acad Sci	&{'weight': 187}\\
	\hline
	Huazhong Univ Sci \& Technol	&Wuhan Univ&	{'weight': 163}\\
	\hline
	Univ Milan	&Fdn IRCCS Ca Granda Osped Maggiore Policlin&	{'weight': 154}\\
	\hline
	Univ Melbourne	&	Monash Univ &{'weight': 120}\\
	\hline
	 Univ Toronto&	Univ Hlth Network&	{'weight': 118}\\
	 \hline
	 Univ Cattolica Sacro Cuore	&Fdn Policlin Univ A Gemelli IRCCS&	{'weight': 114}\\
\hline
Univ Toronto&	Sunnybrook Hlth Sci Ctr	&{'weight': 80}\\
\hline
Univ Oxford	&Imperial Coll London &{'weight': 78}\\
\hline
\end{tabular}
\caption{Institution collaboration pairs}
\label{table:icp}
\end{table}
\pagebreak
\begin{figure}[h!]
    \centering
    \includegraphics[width=\linewidth]{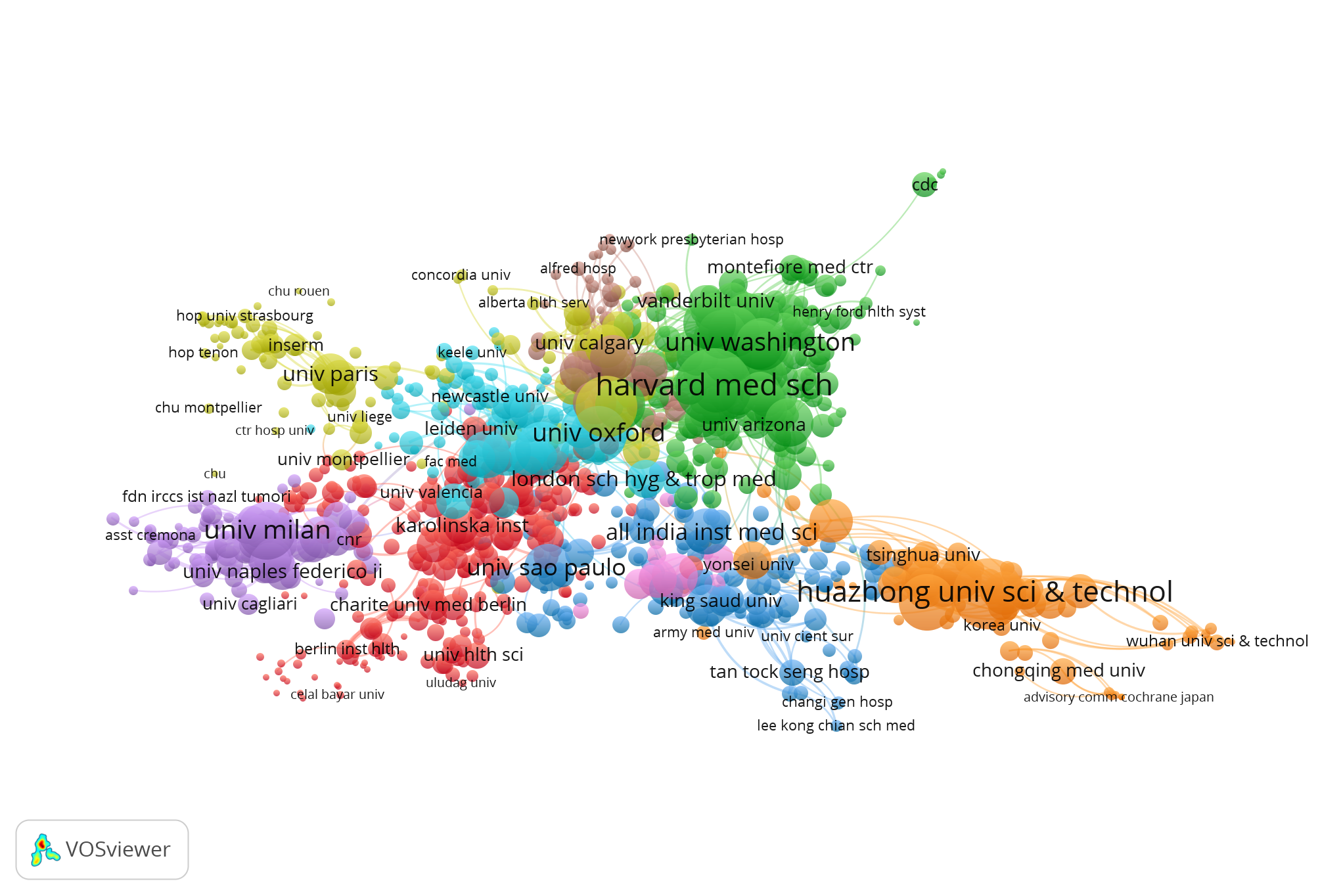}
    \caption{Network Visualization of institution collaboration network}
    \label{fig:instvos}
\end{figure}
\vspace{10cm}
\pagebreak
\subsubsection{Research area co-occurrence network}
A research area co-occurrence network is a network where each node represents a research area, and two nodes are connected by a link if two research areas co-occurrent together in one paper. Lists of research area are from the 'Research Area' column in the original data set. The purpose of building it is to have a general picture, such as which two research areas are most frequently appear together.
\\\\
By applying functions from NetworkX, some facts are calculated.\\Facts of the Research area co-occurrence network:\\
\begin{itemize}
    \item Number of nodes: 151
    \item Number of edges: 1,071
    \item Number of isolated nodes: 1
    \item Number of connected components: 1
    \item Size of the largest component: 150
\end{itemize}
The most frequently appeared research area pairs are shown in the following table. Infectious Diseases and Immunology as complement subjects appear together the most. Infectious Diseases also mostly associate with Public, Environmental \& Occupational Health, Microbiology, etc.
\begin{table}[h!]
\centering
\begin{tabular}{ |p{5cm}|p{5cm}|p{2.5cm}| } 
\hline
\textbf{Research area} &	\textbf{Research area} & \textbf{Count}\\
\hline
Infectious Diseases&	Immunology&	{'weight': 948}\\
\hline
	Infectious Diseases	&Public, Environmental \& Occupational Health	&{'weight': 919}\\
\hline	
	Public, Environmental \& Occupational Health&	Environmental Sciences \& Ecology	&{'weight': 918}\\
	\hline
	Infectious Diseases&	Microbiology&	{'weight': 883}\\
\hline	
	Microbiology	&Immunology&	{'weight': 703}\\
\hline	
	Psychiatry&	Psychology&	{'weight': 614}\\
\hline	
	Biochemistry \& Molecular Biology&	Cell Biology&	{'weight': 592}\\
\hline	
	General \& Internal Medicine&	Respiratory System&	{'weight': 538}\\
\hline	
	Medical Informatics	&Health Care Sciences \& Services&	{'weight': 461}\\
\hline	
	Cardiovascular System \& Cardiology	&Hematology&	{'weight': 425}\\
\hline	
\end{tabular}
\caption{Research area co-occurrence}
\label{table:raco}
\end{table}
\pagebreak
\subsubsection{Keyword co-occurrence network}
A keyword co-occurrence network is a network where each node represents a keyword, two nodes are connected by an edge if two keywords appear in one paper. The keyword data is from the 'Author keyword' column in the original data set. Since the keywords are added by WoS, there is a more systematic keyword dictionary. Similar to what has being done in research area network, the purpose of the keyword co-occurrence network is to find out what author keywords appear together most frequently.\\
\\
By applying functions from NetworkX, some facts were calculated.\\
Facts of the Keyword co-occurrence network:
\begin{itemize}
    \item Number of nodes: 62,565
    \item Number of edges: 435,361
    \item Number of isolated nodes: 18
    \item Number of connected components: 328
    \item Size of the largest component: 61,039
    \item Size of top 5 largest components: 61,039, 19, 17, 13, 12    
\end{itemize}
The most frequently appeared author keyword pairs are shown in the following table. Aside from the equivalent terminologies, COVID-19 also associates with mental health, public health, mortality and anxiety.
\begin{table}[h!]
\centering
\begin{tabular}{ |p{4cm}|p{4cm}|p{4cm}| } 
\hline
\textbf{Author keyword} &	\textbf{Author keyword} & \textbf{Count}\\
\hline
covid-19&	sars-cov-2&	{'weight': 7,106}\\
\hline
	covid-19&	coronavirus&	{'weight': 4,107}\\
\hline	
	covid-19&	pandemic&	{'weight': 3,024}\\
\hline	
	coronavirus&	sars-cov-2&	{'weight': 1,613}\\
\hline	
	coronavirus&	pandemic&	{'weight': 915}\\
\hline	
	covid-19&	mental health&	{'weight': 686}\\
	\hline
	sars-cov-2&	pandemic&	{'weight': 671}\\
	\hline
	covid-19	&public health&	{'weight': 620}\\
	\hline
	covid-19&anxiety	&	{'weight': 567}\\
\hline	
	covid-19&mortality&	{'weight': 564}\\
\hline	
\end{tabular}
\caption{Author keyword co-occurrence}
\label{table:keyco}
\end{table}
\pagebreak
\begin{figure}[h!]
    \centering
    \includegraphics[width=\linewidth]{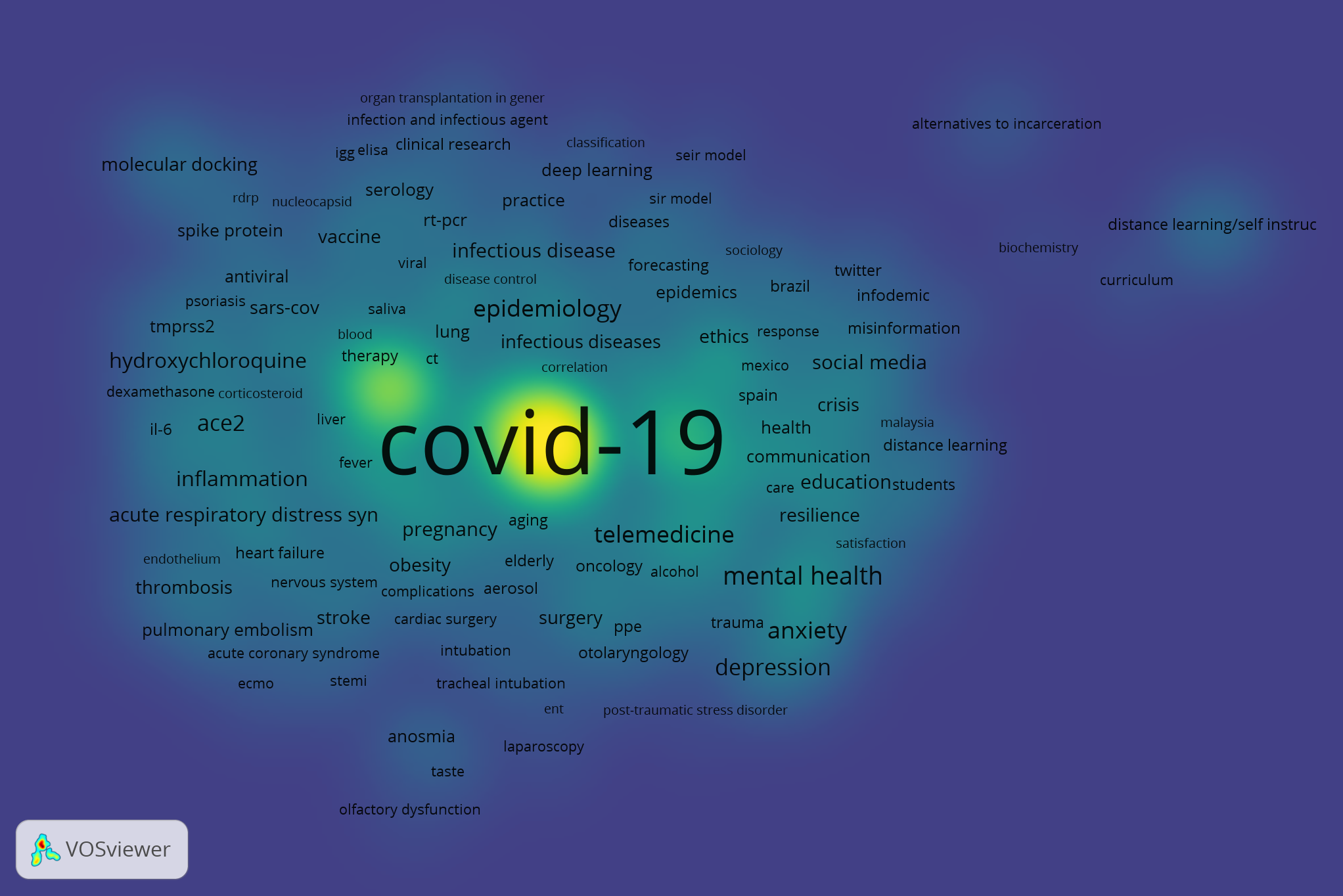}
    \caption{Density Visualization of author keywords}
    \label{fig:keyvos}
\end{figure}
\pagebreak
\clearpage
\newpage
\subsubsection{Centrality}
In network analysis, indicators of centrality give the importance of vertices. The word 'importance' has a wide number of meanings, leading to many different definitions of centrality. Common centralities [19] are degree centrality, betweenness centrality and closeness centrality.\\
\\
The country centrality table above gives a new picture in terms of each country's importance in the collaboration network. Nevertheless, USA still remained at the top.
\begin{table}[h!]
\centering
\begin{tabular}{ |p{3cm}|p{2.8cm}|p{2.8cm}|p{2.8cm}| } 
\hline
\textbf{Country} &	\textbf{betweeness} & \textbf{closeness} & \textbf{degree}\\
\hline
USA&	0.096462&	0.911215 & 0.917949\\
\hline
	France&	0.050390&	0.802469 & 0.758974\\
	\hline
	Australia&	0.038438&	0.776892 & 0.712821\\
	\hline
	UK&	0.034480&	0.836910 & 0.815385\\
	\hline
	Italy&	0.025609&	0.799180 & 0.753846\\
	\hline
	Spain&	0.023060&	0.767717 & 0.702564\\
	\hline
	India&	0.022449&	0.799180 & 0.748718\\
	\hline
	Pakistan&0.021882&	0.764706& 0.692308\\
	\hline
	China&	0.018578&	0.789474 & 0.733333\\
	\hline
	Switzerland&	0.018540&	0.786290& 0.728205\\
\hline	
\end{tabular}
\caption{Country centrality}
\label{table:cenc}
\end{table}
\pagebreak
\subsubsection{Small-world property of co-authorship network}
In this co-authorship network, consider only the largest component that has 189,633 nodes. By calculating with NetworkX function, the average shortest path length is 17.26, and the natural logarithm of the number of nodes is 12.15, since those two numbers have the same order of magnitude, it is indeed a small-world network. Furthermore, the average local clustering coefficient is 0.87, it also supports the small-world property.
\subsubsection{Scale-free property of co-authorship network}
Powerlaw library is imported to Jupyter notebook to verify the scale-free property in the co-authorship network.
\begin{figure}[h!]
    \centering
    \includegraphics[width=0.7\linewidth]{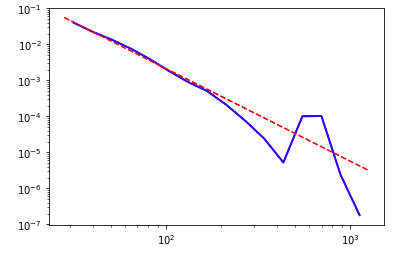}
    \caption{Fitting powerlaw}
    \label{fig:powerlaw}
\end{figure}
\pagebreak
\\
As it shows, the degree distribution asymptotically follows the power law distribution. Especially the first half fits the powerlaw almost perfectly. So the co-authorship network can be considered as a scale-free network. The degree sequence and their frequency of appearing can be shown in the following:\\
\begin{figure}[h!]
    \centering
    \includegraphics[width=0.7\linewidth]{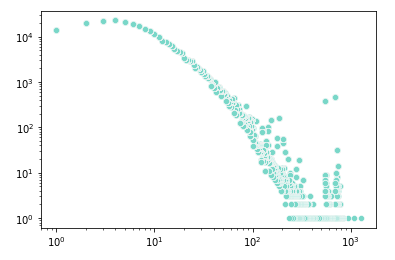}
    \caption{Degree sequence}
    \label{fig:deg}
\end{figure}
\pagebreak
\subsubsection{Assortativity}
In co-authorship network, assortativity coefficient is calculated for each component sorted in size decreasing order. However, the assortativity coefficient for regular graph is not well-defined because the denominator will be zero in such cases, and it will give a nan result. Hence, the nan results are discarded from the sequence. The following graph shows the change of assortativity coefficient as the size of the component decreases. \\
\begin{figure}[h!]
    \centering
    \includegraphics[width=0.7\linewidth]{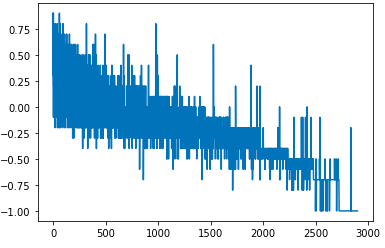}
    \caption{Assortativity change}
    \label{fig:assort}
\end{figure}
\pagebreak
\\As it suggests, there is a decreasing trend in the assortativity coefficient. The bigger the size of the component, the more correlated the two nodes are. There are cases where in one component, nodes are perfectly uncorrelated, and those appear in components with small sizes. Although the components with perfectly correlated nodes are also detected, but they only appear among the large components, and they are rather rare.
\section{Conclusion}
In 2020, the cumulative research output related to COVID-19 was continuously rising. The quantitative output in most aspects rose significantly in the first half of the year, while remained stagnent in the second half, but still stayed at a high level. Top research countries such as USA, China, and UK, played a significant part in the COVID-19 research, while domestic collaboration remained popular throughout the year. Many publication sources are found, with BMJ-British Medical Journal as the leading publication source. General \& Internal Medicine was the top research area, which was also most top cited papers' research topic. COVID-19 related publications are published mostly as collaborated papers, and mostly characterized by only one research area. There is also a great portion of COVID-19 related publications that studies the psychological impact rather than Medicine. The co-authorship network has small-world and scale-free property, which suggests that the collaborators of any author are likely to be collaborators of each other and most authors can be reached by a small number of steps.\\
\\
Similar to the majority of the studies, a general growing trend of the number of the publications is observed, due to the efforts made by scientific researchers. Meanwhile, measures should also be taken to encourage future study in this field.

\end{document}